\documentclass[twocolumn]{article}
\usepackage{natbib}

\title{\bf Magnetic Fields of Black Holes and the Variability Plane}

\author{M.Yu. Piotrovich, N.A. Silant'ev,
 Yu.N. Gnedin\thanks{E-mail: gnedin@gao.spb.ru},
 T.M. Natsvlishvili\\ Central Astronomical Observatory at
 Pulkovo, 196140, Saint-Petersburg, Russia.}

\begin{document}

\maketitle

\begin{abstract}
We estimated the magnetic field strength at the horizon radius of
black holes, that is derived by the magnetic coupling process and
depended on the black hole mass $M_{BH}$ and the accretion rate
$\dot{M}$. Our estimation is based on the use of the fundamental
variability plane for stellar mass black holes, AGNs and QSOs. The
typical values of magnetic field strength on the black hole
horizon are appeared at the level of $B_{BH}\sim 10^8$G for
stellar mass black holes and $B_{BH}\sim 10^4$G for the
supermassive black holes. We have obtained the relation $p_l\sim
\nu^{-1/2}_b$ between the intrinsic polarization of the accretion
disk radiation and the characteristic frequency of the black hole
X-ray variability.
\end{abstract}

\section{Introduction}

Recently, the strong evidence has been obtained that active
galactic nuclei and soft-state black hole X-ray binaries populate
a plane in the space defined by the black hole mass, accretion
rate and characteristic frequency of their X-ray variability
\citep{mchardy06, kording07, casella08}. It should be noted that
there exist and other relations between the basic physical
parameters of central engines supported the general idea about
similarity between X-ray binaries (XRBs) and active galactic
nuclei (AGNs). First, the, so-called, fundamental plane of
accreting black holes connects XRBs and AGNs through a plane in
the black hole mass, radio and X-ray luminosity space
\citep{merloni03, falcke04}. The space defined by the accretion
rate, the black hole mass and a characteristic time scale of the
X-ray variability suggests that all black holes can be unified by
taking the accretion rate as well as the black hole mass into
account.

The new idea developed by us in this paper is to show that the
magnetic field strength at the event of a black hole horizon lies
in the base of this correlation. We shall show below that above
mentioned correlation allows us to derive directly the magnetic
fields strength on the horizon of a black hole through the
characteristic frequency of the X-ray variability. This fact
allows also to determine the magnetic field strength for a number
of well-known AGNs and XRBs and to compare obtained magnitudes
with corresponding values derived by indirect methods. One of such
methods is observation of optical polarization of radiation from
an accreting disk around the central black holes. This method
based on taking into account the effect of Faraday rotation of
polarization plane at multiple scattering of accretion disk
radiation has been developed in series of papers by Gnedin and
Silant'ev and their co-authors \citep{gnedin84, gnedin97,
dolginov95, gnedin06, silantev02, silantev05} and also in papers
\citep{agol96, agol98, shternin03}.

The power spectral density (PSD) is usually described by a number
of Lorentzians \citep{psaltis99, belloni05}. Each Lorentzian is
described by their characteristic frequency, i.e. the frequency of
the maximum of the Lorentzian in the frequency times power plot.

The measured shapes of AGN PSDs are similar to those of soft-state
stellar black holes, but time scales for AGNs are several orders
of magnitudes longer. It means that characteristic frequency
$\nu_B$ for AGNs is, as a rule, less that its value in the case of
stellar black holes. Nevertheless, the dependence of the
variability properties on accretion rate and mass of a black hole
takes similar forms both for stellar black holes, and for AGN.

\section{The Variability Plane}

\citet{mchardy06} have shown that galactic black holes in their
soft-state and AGNs populate a plane in the parameter space of the
black hole mass, accretion rate and characteristic frequency in
PSD. \citet{kording07} recently presented evidence for this
variability plane to extend to the hard-states BHs. At last,
\citet{casella08} extended this plane to the black holes in
ultraluminous X-ray sources. For two ULXs M82 X-1 and NGC 5408 X-1
they estimate a black hole mass in the range $100\div 1300
M_{\odot}$. This result gives a definite evidence that ULX sources
may be considered as intermediate mass black holes (IMBH).

The variability plane can be presented in the following form
\citep{kording07, casella08}:

\begin{equation}
 \log{\nu_b} = \log{\dot{M}} - 2 \log{M} - 14.7 - 0.9\theta,
 \label{eq1}
\end{equation}

\noindent where $\dot{M}$ is the accretion rate in g/s, $M =
M_{BH} / M_{\odot}$ (here and below) is the black hole mass in
$M_{\odot}$ and $\theta$ goes from 0 for the soft-state stellar
black holes to 1 for hard-state BHs, $\nu_b$ is the break
frequency in Hz. For AGNs the value $\theta = 0$ can be accepted.

The magnetic field strength is derived in a result of the magnetic
coupling (MC) process. We remind that the MC process is
responsible for interaction between a rotating black hole and its
surrounding accretion disk. This process is commonly accepted as a
variant of the famous Blandford and Znajek (BZ) process
\citep{blandford77, li02, wang02, wang03, aloiso04, uzdensky05,
zhang05, gan09}. In this process energy and angular momentum are
transferred from a rotating BH to its surrounding disk. It can
also depress accretion via transfer of angular momentum from a
rotating BH to the inner disk. It is important that MC is
responsible for generating the outflows and jets from the BH
binaries and AGNs.

The MC process provides the relation between the magnetic field
strength at the horizon of BH $B_H$ and its mass $M$ and the
accretion rate $\dot{M}$. This relation is based upon the balance
between the pressure of the magnetic field at the horizon and the
accreted matter pressure of the innermost part of an accretion
flow:

\begin{equation}
 B_H = \frac{1}{R_H} \sqrt{2 k_m \dot{M} c},
 \label{eq2}
\end{equation}

\noindent where the radius of the BH horizon $R_H$ is

\begin{equation}
 R_H = \frac{G M_{\odot}}{c^2} M \left[1 + \sqrt{1 -
 a_*^2}\right].
 \label{eq3}
\end{equation}

\noindent Here $a_*$ is the dimensionless Kerr parameter (spin).
For Schwarzshild BH $a_* = 0$. $k_m$ is a magnetization parameter
indicating the relative power of the MC process with respect to
the disk accretion. If the MC process dominates over the disk
accretion, $k_m > 1$, and if it is dominated by latter, $k_m < 1$.
The case $k_m \approx 1$ corresponds to equipartition between
these two processes.

At logarithm scaling, Eq.(\ref{eq2}) corresponds to

\begin{equation}
 2 \log{B_H} = \log{\dot{M}} - 2 \log{M} + 2 \log{f(a_*,k_m)},
 \label{eq4}
\end{equation}

\noindent where

\begin{equation}
 f(a_*,k_m) = \frac{c^2 \sqrt{2 k_m c}}
 {G M_{\odot} \left[1 + \sqrt{1 - a_*^2}\right]} \cong
 \frac{1.66 \sqrt{k_m}}{1 + \sqrt{1 - a_*^2}}.
 \label{eq5}
\end{equation}

Eqs.(\ref{eq1}) and (\ref{eq4}) gave together:

\begin{equation}
 B_H = \left[\nu_{b} f^2(a_*,k_m)\right]^{1/2} \times
 10^{7.35 + 0.45 \theta}.
 \label{eq6}
\end{equation}

\noindent Here we shall use the value $\theta \approx 0$
\citep{mchardy06}.

\section{Magnetic Field Strength of Stellar Black Hole from their
X-ray Timing}

We are starting with estimates of magnetic field strength at the
horizon of black hole from X-ray binaries.

Recently \citet{mchardy06} have demonstrated that the break
frequency $\nu_b$ is determined by both the black hole mass
$M_{BH}$ and the accretion rate $\dot{m}_{Edd} = \dot{M} /
\dot{M}_{Edd}$ as follows:

\begin{equation}
 \nu_b = 0.029 \varepsilon \dot{m}_{Edd} (10^6 M_{\odot} /
 M_{BH}),
 \label{eq7}
\end{equation}

\noindent where $\varepsilon$ is the efficiency of the mass to
energy conversion (usually one assumes $\varepsilon = 0.1$). For
Cyg X-1 with $\dot{m}_{Edd} = 0.1$ and $M_{BH} = 10 M_{\odot}$ the
value $\nu_b = 29$Hz and $B_H \approx 1.2 \times 10^8$G. This
magnetic field magnitude corresponds quite well to the results of
polarimetric observations \citep{gnedin06, karitskaya09a,
karitskaya09b}. For other stellar mass black holes the typical
value of the break frequency lies into the interval $160 \div
300$Hz \citep{tomsick09}. Then, using the estimate by
Eq.(\ref{eq6}), we obtain the following value of the magnetic
field strength at the horizon radius of Kerr black hole: $B_H
\approx 4 \times 10^8$G for the equipartition condition, i.e. $k_m
= 1$. The numerical calculations for the specific stellar mass
black holes are presented in Table 1.

Recently \citet{mcclintock09} have fitted the thermal continuum
X-ray spectra of black hole X-ray binaries using the
Novikov-Thorne thin disk model. As a result they have extracted
the dimensionless spin parameter $a_*$ for some X-ray binary black
holes. They have obtained $a_* = 0.65 \div 0.75$ for J1665-40 and
$a_* = 0.98 \div 1$ for GRS 1915+105. Supposing $a_* = 0.7$, we
obtain the following estimation of magnetic field $B_H = 3 \times
10^8$G for J1665-40.

\begin{table}
 \caption{The results of our calculations ($k_m = 1$).}
 \label{tab1}
 \begin{tabular}{@{}lcrr}
 \hline
 Source        & $\nu_b(Hz)$ & $B_H$(G)          & $B_H$(G) \\
               &                 & ($a_* = 0$)       & ($a_* = 0.998$) \\
 \hline
 GRO J1665-40  & 300             & $3.2\times 10^8$  & $6.4\times 10^8$ \\
 H 1743-322    & 241             & $2.9\times 10^8$  & $5.8\times 10^8$ \\
 XTE 1550-64   & 276             & $3.1\times 10^8$  & $6.2\times 10^8$ \\
 GRS 1915+105  & 168             & $2.4\times 10^8$  & $4.8\times 10^8$ \\
 4U 1630-47    & 164             & $2.38\times 10^8$ & $4.76\times 10^8$ \\
 XTE J1650-500 & 250             & $2.9\times 10^8$  & $5.8\times 10^8$ \\
 XTE 1859+228  & 190             & $2.6\times 10^8$  & $5.2\times 10^8$ \\
 \hline
\end{tabular}
\end{table}

\section{Magnetic Field Strength of AGNs from their Quasiperiodic
Oscillations}

For SMBH (AGNs) the typical values of the break frequency lie in
the ranges $\nu_b = 10^{-5} \div 10^{-7}$Hz. It allows to estimate
the typical magnetic field strength for Kerr black hole ($a_* =
0.998$) $B_H \approx 10^4 \div 10^5$G.

Let us consider now the specific targets.

For the radio-loud active galactic nucleus 3C 390.3 the break
frequency values is $\nu_b = 2.7 \times 10^{-7}$
\citep{gliozzi09}. For the horizon magnetic field strength one
obtains $B_H = 10^4$G for the Schwarzschild black hole ($a_* = 0$)
and $B_H = 2 \times 10^4$G for the Kerr type black hole. PSD
studies of the blazars Mrk 421, Mrk 501 and PKS 2155-304 with
jet-dominated accretion disks suggest the presence of PSD breaks
at frequencies that are nearly 2 orders of magnitude higher that
the break found in 3C 390.3 i.e. at $\nu_b \approx 10^{-5}$Hz
\citep{gliozzi09}. For these targets the horizon magnetic field
strength $B_H = 5.8 \times 10^4$G in the case of the Schwarzschild
black hole and is twice higher for the Kerr black hole with $a_*
\approx1$.

Interesting case of sources is the AGN SDSS J0013-0951. The
physical parameters of this target $M_{BH} = 8.09 \times 10^5
M_{\odot}$, $\dot{m}_{Edd} = 0.4051$ \citep{lamura07}. The break
frequency for this AGN can be estimated from the Eq.(\ref{eq7}).
In this case $\nu_b = 0.015\varepsilon$ and the horizon magnetic
field strength is determined by this relation:

\begin{equation}
 B_H = 4.45 \times 10^6
 \frac{\sqrt{k_m \varepsilon}}{1 + \sqrt{1 - a_*^2}}.
 \label{eq8}
\end{equation}

For the Schwarzschild type of black hole $B_H = 5.3 \times 10^5
\sqrt{k_m}$G, for the Kerr type of black hole $B_H = 2.5 \times
10^6 \sqrt{k_m}$G. For Mrk 766 $\nu_b = 5 \times 10^{-4}$Hz
\citep{papadakis09} and $B_H = 4 \times 10^5 \sqrt{k_m}$G ($a_* =
0$), $B_H = 8 \times 10^5 \sqrt{k_m}$G ($a_* = 0.998$).

The results of our calculations for a numbers of AGN and QSO are
presented in Table 2. The data accepted here for $\nu_b$ are taken
from \citet{papadakis09}.

\begin{table}
 \caption{The results of our calculations ($k_m = 1$).}
 \label{tab2}
 \begin{tabular}{@{}lrrr}
 \hline
 Source       & $\nu_{b}(Hz)$       & $B_H$(G)          & $B_H$(G) \\
              &                      & ($a_* = 0$)       & ($a_* = 0.998$) \\
 \hline
 3C 390.3     & $2.7\times 10^{-7}$  & $10^4$            & $2\times 10^4$ \\
 Mrk 421      & $\sim 10^{-5}$       & $5.8\times 10^4$  & $1.2\times 10^5$ \\
 Mrk 501      & $\sim 10^{-5}$       & $5.8\times 10^4$  & $1.2\times 10^5$ \\
 PKS 2155-304 & $\sim 10^{-5}$       & $5.8\times 10^4$  & $1.2\times 10^5$ \\
 NGC 5548     & $6.3\times 10^{-7}$  & $1.5\times 10^4$  & $2.8\times 10^4$ \\
 Ark 564      & $2.3\times 10^{-3}$  & $9\times 10^5$    & $1.7\times 10^6$ \\
 Mrk 766      & $6.1\times 10^{-4}$  & $4.55\times 10^5$ & $8.6\times 10^5$ \\
 NGC 4051     & $5.05\times 10^{-4}$ & $4.23\times 10^5$ & $8\times 10^5$ \\
 Fairall 9    & $4\times 10^{-7}$    & $1.88\times 10^4$ & $3.7\times 10^4$ \\
 PG 0804+761  & $9.6\times 10^{-7}$  & $1.8\times 10^4$  & $3.6\times 10^4$ \\
 NGC 3227     & $2\times 10^{-5}$    & $8.3\times 10^3$  & $1.66\times 10^4$ \\
 NGC 3516     & $2\times 10^{-6}$    & $2.6\times 10^4$  & $4.95\times 10^4$ \\
 NGC 3783     & $4\times 10^{-6}$    & $3.7\times 10^4$  & $7\times 10^4$ \\
 NGC 4151     & $1.3\times 10^{-6}$  & $2.12\times 10^3$ & $4\times 10^3$ \\
 NGC 4258     & $2\times 10^{-8}$    & $2.6\times 10^3$  & $5\times 10^3$ \\
 MCG-6-30-15  & $7.7\times 10^{-5}$  & $1.6\times 10^5$  & $3\times 10^5$ \\
 NGC 5506     & $1.3\times 10^{-5}$  & $6.6\times 10^4$  & $1.2\times 10^5$ \\
 RE J1034+396 & $2.5\times 10^{-4}$  & $2.9\times 10^5$  & $4.56\times 10^5$ \\
 \hline
\end{tabular}
\end{table}

\section{Relation between Linear Polarization and Break
Frequency}

Magnetic field in an accretion disk has strong influence on the
accretion disk polarization due to the Faraday rotation effect of
the polarization plane on the free path length of a photon in the
accretion disk \citep{dolginov95, gnedin97, silantev02, agol96,
gnedin06}. \citet{silantev09} have introduced the relative
polarization degree that is

\begin{equation}
 p_{rel} = \frac{p_l ({\bf B}, \mu)}{p_l (0, \mu)} =
 \frac{1}{[1 + 2 (a^2 + b^2) + (a^2 - b^2)^2]^{1/4}},
 \label{eq9}
\end{equation}

\begin{equation}
 \tan{2 \chi} = \frac{2 a}{(1 / p_{rel})^2 + 1 + b^2 - a^2}.
 \label{eq10}
\end{equation}

\noindent Here, $p_l(0,\mu)$ is the known Sobolev-Chandrasekhar
polarization degree in the Milne problem \citep{chandr50}, $\chi$
is the position angle ($\chi = 0$ denotes oscillations to the disk
plane).

The parameters of the Faraday depolarization $a$ and $b$ are
connected with the components $B_z$ and $B_{\bot}$ of the global
magnetic field in the accretion disk:

\[
 a = 0.8 \left(\frac{\lambda_{rest}}{1\mu m}\right)^2
 \left(\frac{B_z}{1 G}\right) \mu,
\]
\begin{equation}
 b = 0.8 \left(\frac{\lambda_{rest}}{1\mu m}\right)^2
 \left(\frac{B_{\bot}}{1 G}\right) \sqrt{1 - \mu^2}.
 \label{eq12}
\end{equation}

\noindent Here $B_z$ is the component directed along the normal to
the accretion disk surface, $B_{\bot}$ is the superposition of the
azimuthal and radial component of the magnetic field in the
accretion disk. $\mu = \cos{i}$, where $i$ is the inclination
angle.

Eqs.(\ref{eq9})-(\ref{eq12}) allow to get the relation between
$\nu_b$ and polarization parameters $P_l$ and $\chi$. Using these
Eqs. and Eq.(\ref{eq6}), one can obtain the following relations in
the asymptotical case when the depolarization parameters $a, b \gg
1$:

\begin{equation}
 P_l(B, \mu) \sim 1 / \sqrt{\nu_b},\,\, \tan{2 \chi} \sim
 \sqrt{\nu_b}\,\,\, for\,\, B_z \gg B_{\bot}
 \label{eq13}
\end{equation}

\begin{equation}
 P_l(B, \mu) \sim 1 / \sqrt{\nu_b},\,\, \chi \approx 0
 \,\,\, for\,\, B_{\bot} \gg B_z
 \label{eq14}
\end{equation}

\section{Conclusions}

We show that the plane in the space defined by the black hole
mass, accretion rate and characteristic frequency of their X-ray
variability, which is populated by active galactic nuclei and
soft-state black hole X-ray binaries, allows to estimate the
magnetic field strength near the event horizon of black holes. The
typical values of magnetic field strength at the black hole
horizon are $\sim 10^8$G for the stellar mass black holes and
$\sim 10^4$G for the supermassive black holes. The value $B_H \sim
10^8$G does not contradict estimates made by \citet{robertson03}
from extremely other point of view. They claimed existence instead
of black holes the original type of objects which were called
Magnetospheric Eternally Collapsing Objects (MECO). They predicted
the existence of intrinsic magnetic moments of $\sim 4 \times
10^{29} G cm^3$ in these objects.

The magnetic field magnitudes estimated for the stellar mass black
holes correspond quite well to the observation data
\citep{karitskaya09a,karitskaya09b}.

\section*{Acknowledgements}

This research was supported by the RFBR (project No.
07-02-00535a), the program of Prezidium of RAS ''Origin and
Evolution of Stars and Galaxies'', the program of the Department
of Physical Sciences of RAS ''Extended Objects in the Universe''
and by the Grant from President of the Russian Federation ''The
Basic Scientific Schools'' NS-6110.2008.2. M.Yu. Piotrovich
acknowledges the Council of Grants of the President of the Russian
Federation for Young Scientists, grant No. 4101.2008.2.

\end{document}